\documentclass[12pt]{article}

\input epsf

\def\i{\mbox{i}}
\def\d{\mbox{d}}
\def\Im{\mbox{Im}}
\def\Re{\mbox{Re}}

\title{ Modified perturbation theory approach
for \\ $t \bar{t}$ production and decay }
\author{M. L. Nekrasov \\
{\small\it Institute for High Energy Physics, 142284 Protvino,
Russia}}

\date{}

\begin{document}

\maketitle

\begin{abstract}

\noindent The modified perturbation theory (MPT), based on direct
expansion of probabilities instead of amplitudes, allows one to
avoid divergences in the phase-space integrals resulting from
production and decay of unstable particles. In the present paper
the range of applicability of MPT is determined numerically in the
case of the process $e^+e^- \to (\gamma,Z) \to t\bar t \to
W^{+}b\:W^{-}\bar b$. It is shown~that with the complete expansion
in powers of the coupling constant (without Dyson resummation) MPT
operates best at the energies located near the maximum of the
cross-section and slightly above the maximum. In this region the
MPT expansion within the next-to-leading order considerably
exceeds in accuracy well-known DPA approach.

\end{abstract}

\section{Introduction}

The problem of high-precision description of the processes with
production and decay of unstable particles is consistently of
great importance, in particular, in view of the forthcoming time
of electron linear colliders (TESLA, NLC, CLIC). The solving of
this problem requires methods maintaining gauge invariance. One
such method is the pole scheme \cite{Pole} based on expansion of
the amplitude in Laurent series around the mass-shell of unstable
particles. In the case of pair production of unstable particles
the leading order (LO) of this expansion is known as the double
pole approximation (DPA). In the case of W-boson pair production
studied at LEP2 the DPA was the sole approach where the one-loop
radiative corrections had been actually evaluated \cite{4f}.

Nevertheless, the DPA itself does not provide a high enough
precision. At LEP2, while calculating the one-loop corrections,
this difficulty was concealed by the additional small factor
$\alpha/\pi$ which diminished the systematic error to the
acceptable level. However in the Born approximation the systematic
error was too large, such large that it forced one to use the
gauge-dependent CC03 off-shell cross-section instead of the DPA
Born cross-section required conceptually and for preserving
gauge-invariance. As a whole the situation was considered
acceptable for the needs of LEP2 \cite{4f}. However it hardly
probable can be considered acceptable, both at the Born and the
one-loop level, in the case of the high-precision calculations
needed for the experiments on future linear colliders \cite{LC}.

Fortunately the pole scheme is not the sole approach that
maintains the gauge-invariance. This property should be inherent
in the modified perturbation theory (MPT) \cite{FT,N1,N2}, as
well, which is based on direct expansion of probabilities in
powers of the coupling constant instead of amplitudes. The
expansion of probabilities in MPT is made possible due to the
applying of the methods of distributions \cite{Distr} which allow
one to escape the divergences in the phase-space integrals. As a
result the probability in MPT is represented as an expansion in a
power series in the coupling constant. This means that one can
expect the gauge invariance of each term of the expansion, at
least, if to believe in gauge invariance of the exact result.

Unfortunately, the above picture not everywhere is completely
legitimate since the MPT series not everywhere is well-defined and
not everywhere has admissible convergence properties. In fact, the
MPT series is ill-defined when the energy takes a value of the
threshold of the production of unstable-particles \cite{B}. (It is
worth emphasizing the difference between the threshold and the
mass shell in the case of the pair, or multiple production of
unstable particles.) One way to solve this problem is to partially
resum the MPT series, by the analogy with the Dyson resummation in
the amplitude. The result of such resummation can be different
from that in the conventional approach where the Dyson resummation
is carried out in the amplitude and only then the probability is
calculated (see discussion in \cite{N1,N2}). Nevertheless, the
gauge invariance in MPT after the Dyson resummation all the same
can be broken, and so the question about the gauge cancellations
in each case should be additionally examined. This problem should
not arise in the MPT series with the complete expansion in the
coupling constant. So, it is important to find a range of energies
where the MPT series well converges. If there is such a range and
if the MPT is able to give there more precise description than the
DPA approach, then the MPT is preferable to carry out the
high-precision calculations in this range of energies.

The searching for this range of energies is the main aim of the
present paper. The study is carried out on an example of the
process $e^+e^- \to (\gamma,Z) \to t\bar t \to W^{+}b\:W^{-}\bar
b$ which is a part of the more general process $e^+ e^- \to 6f$
planed for future linear colliders. In the present paper the
above-mentioned process is considered in the Born approximation
for the cross-section but with the taking into consideration of
the one-loop and two-loop corrections in the Dyson resummed
propagators before applying the MPT expansion. This is the
necessary first step for a key solution of all the problem. By the
subsequent step the similar examination should be carried out in
more realistic case which would more precisely simulate the
complete process $e^+ e^- \to 6f$ with taking into consideration
the other loop corrections. Some aspects of this consideration
concerning, in particular, the non-factorizable corrections, have
already been discussed in \cite{N1,N2}.

The structure of the present paper is as follows. In the next
Section we discuss the details of the definition of the Born
cross-section for the process $e^+ e^- \to (\gamma, Z) \to t\bar t
\to W^{+} b\:W^{-}\bar b$ with the Dyson resummation in the
$t$-quark propagators. In Section~3 the same process is considered
in MPT approach within the next-to-leading-order (NLO)
approximation. In Section~4 we present the results of numerical
calculations and compare the ``exact'' result of Section~2 with
the results of MPT and DPA approaches. Section~5 summarizes the
main results of the paper.

\section{The statement of the problem}

\begin{figure}[t]
\[
\epsfbox{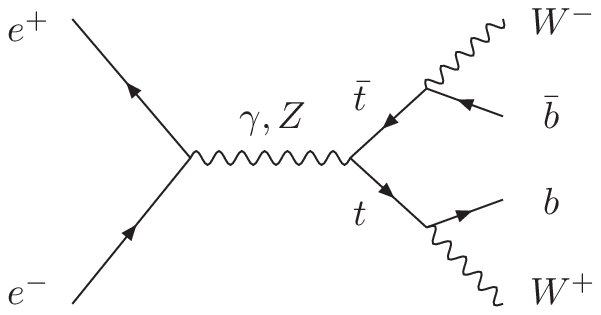}
\]
\caption{The process $e^+e^- \to (\gamma,Z) \to t\bar t \to
W^{+}b\:W^{-}\bar b$.}
\end{figure}

The total cross-section of the unpolarized process $e^+ e^- \to
(\gamma, Z) \to t\bar t \to W^{+} b\:W^{-}\bar b$ (see Fig.~1)
under the supposition of stability of $W$ and $b$, with massive
$W$ bosons and massless $b$ quarks reads as follows:
\begin{equation}\label{1}
 \sigma (s) =
 \int\limits_{4 M_W^2/s}^1 \d z \> \phi(z;s) \> \hat\sigma(zs)\,,
\end{equation}
\begin{equation}\label{2}
\hat\sigma (s) = \int\limits_{M_W^2}^{(\sqrt{s}-M_W)^2}\!\!\!\!\d
s_+ \int\limits_{M_W^2}^{(\sqrt s\;- \sqrt {s_+})^2} \!\!\!\!\d
s_- \>\hat\sigma(s\,;s_+,s_-) .
\end{equation}
Here $\phi (z;s)$ is the flux function describing the large QED
corrections that arises due to virtual and real photon radiation
from initial states,
\begin{eqnarray}\label{3}
&\phi (z;s) = \beta_e (1-z)^{(\beta_e -1)}
             - \frac{1}{2} \beta_e (1+z), \qquad
\beta_e = \frac{2 \alpha }{\pi}\left(\ln \frac{s}{m_e^2}
-1\right)\,.&
\end{eqnarray}
Variables $s_{\pm}$ stand for the invariant masses of the
final-state systems $W^{+}b$ and $W^{-}\bar b$. The quantity
$\hat\sigma(s\,;s_+,s_-)$ is the exclusive cross-section of the
off-shell process with the Dyson resummation in $t$-quark
propagators. In general case $\hat\sigma(s\,;s_+,s_-)$ can be
written as
\begin{equation}\label{4}
\hat\sigma(s\,;s_+,s_-) =
      \frac{1}{s^2}\sqrt{\lambda (s;s_{+},s_{-})}\;F(s;s_+,s_-)
      \> \rho (s_{+}) \> \rho (s_{-})\,.
\end{equation}
In formula (4) we have explicitly extracted the density of flow of
the colliding particles and the $t \bar t$ phase-volume factor.
Function $F$ is to be determined by direct calculations in the
conventional perturbation theory with taking into account the
Breit-Wigner factors which are chosen in the form
\begin{equation}\label{5}
\rho (s) = \frac{M \Gamma_0}{\pi}\times |\Delta(s)|^2\,.
\end{equation}
Here $M$ is the mass of $t$ quark, $\Gamma_0$ is its Born width,
and $\Delta(s)$ is the scalar factor of the Dyson resummed
propagator of $t$ quark. In general case we have
\begin{equation}\label{6}
\Delta(s) =
 \frac{1}{s - M^2 + \Re \Sigma(s) + \i\:\Im \Sigma(s)}\,,
\end{equation}
where $\Sigma(s)$ is the scalarized self-energy. We suppose that
$\Sigma(s)$ is determined in the on-mass-shell (OMS) scheme of UV
renormalization, more precisely, in some variant of its
generalization to the unstable-particle case. Actually there are
many such variants. The conventional one \cite{OMS} implies the
gauge-dependence in the renormalized mass at the two-loop order.
The variants free from this defect are considered in
\cite{Hollik,OMS-bar}.

It is well-known that the self-energy in the denominator of a
Dyson resummed propagator has to be determined with the exceeding
precision with respect to that of the cross-section. For example,
if $\hat\sigma(s\,;s_+,s_-)$ is to be determined within the
one-loop precision, then $\Sigma(s)$ in (6) has to be determined
at least within the two-loop precision, etc. This rule
follows~from the relation $s - M^2 = O(\alpha)$ which takes place
in the resonance region, so that $\Sigma(s)$ becomes comparable
with $s - M^2$. Outside the resonance region the above-mentioned
rule is irrelevant.

Further on we are interested in the Born approximation for
$\hat\sigma(s\,;s_+,s_-)$, determined with the Dyson resummation
implemented within the NLO precision. This particular quantity is
a part of the complete cross-section determined with NLO
precision. So, we have to define $\Sigma(s)$ in propagator (6) in
the resonance region within the two-loop precision and outside the
resonance region within the one-loop precision.

In respect to the imaginary part of $\Sigma(s)$ the above
requirement can be satisfied by putting
\begin{equation}\label{7}
\Im\Sigma(s) = \sqrt{s}\;\Gamma(s),
\end{equation}
where $\Gamma(s)$ is the width determined within the one-loop
precision (two-loop precision for $\Im\Sigma$) and multiplied by
the Born $s$-dependent factor. In other words, by definition we
put
\begin{equation}\label{8}
\Gamma(s) = \Gamma(M^2) \times f(s)/f(M^2)\,,
\end{equation}
where $f(s) = \sqrt{s\,\lambda(s;M_W^2,0)} \times \left(1 +
M_W^2/s - 2M_W^4/s^2\right)$ \cite{Khoze}, and $\Gamma(M^2)$ is
the sum of the Born and the one-loop contributions to the on-shell
width,
\begin{equation}\label{9}
\Gamma(M^2) = \Gamma_0 +\alpha \Gamma_1 .
\end{equation}
Notice that the self-energy within the same precision can be
written as $\Sigma(s) = \alpha \Sigma_1(s) + \alpha^2\Sigma_2(s)$
where $\alpha\Sigma_1$ and $\alpha^2\Sigma_2$ are the one- and
two-loop contributions, respectively.

The real part of the renormalized self-energy in the most general
case can be written as
\begin{eqnarray}\label{10}
&\Re\Sigma(s) = r^{(0)} + (s-M^2)\,r^{(1)} +
\frac{1}{2}(s-M^2)^2\,r^{(2)}(s) .&
\end{eqnarray}
Here $r^{(0)}$ and $r^{(1)}$ are numerical coefficients, and
$r^{(2)}(s)$ is a function finite at $s=M^2$. The coefficients
$r^{(0)}$ and $r^{(1)}$ are dependent on the UV renormalization
scheme. In the framework of conventional generalization of the OMS
scheme \cite{OMS} they are zero,
\begin{equation}\label{11}
r^{(0)}= 0\,,\qquad r^{(1)}= 0\,.
\end{equation}
In the generalized by \cite{Hollik} OMS scheme they are equal to
\begin{equation}\label{12}
r^{(0)}= -\alpha^2\,\Im\Sigma_1(M^2)\,\Im\Sigma'_1(M^2) +
O(\alpha^3)\,,\qquad r^{(1)}= 0\,.
\end{equation}
In the so called $\overline{\mbox{OMS}}$ scheme \cite{OMS-bar}
they are
\begin{eqnarray}\label{13}
\matrix{ \quad r^{(0)}=
-\alpha^2\,\Im\Sigma_1(M^2)\,\Im\Sigma'_1(M^2)+
O(\alpha^3)\,,\nonumber\\[-0.5\baselineskip]\hfill\cr \quad \>\>
r^{(1)}= -\frac{ 1}{2}
\alpha^2\,\Im\Sigma_1(M^2)\,\Im\Sigma''_1(M^2)+
O(\alpha^3)\,.\hfill\cr}
\end{eqnarray}

Formulas (7)--(13) completely determine the structure of the Dyson
resummed propagator. In what follows we put $r^{(2)}=0$ which is
reasonable in the context of verification of the MPT approach in
the LO+NLO approximation. For determinancy, we adhere to the
$\overline{\mbox{OMS}}$ scheme of UV renormalization.

\section{The LO and NLO in MPT }

The essence of the MPT approach \cite{FT} factually consists in
expanding in the sense of distributions \cite{Distr} of the
propagator squared of unstable particle $|\Delta(s)|^2$ into an
asymptotic series in powers of the coupling constant. Up to
$O(\alpha)$ corrections this series looks as follows:
\begin{equation}\label{14}
|\Delta(s)|^2 = \frac{\pi}{\alpha\,\Im\Sigma_1(M^2)} \,
\delta(s-M^2) - \pi\frac{\Im\Sigma_2(M^2)}{[\Im\Sigma_1(M^2)]^2}\,
\delta(s-M^2)  + V\!P \frac{1}{(s-M^2)^2} + O(\alpha)\;.
\end{equation}
Here $\delta(\cdots)$ is the $\delta$-function, $V\!P$ means the
principal-value prescription. Each term of expansion (14) is to be
integrated with some weight function when substituting into the
formulas for the cross-section. The first term in (14) in view of
(7) represents the well-known formula for the Breigt-Wigner
resonance in the narrow-width approximation. The second and the
third terms describe the NLO corrections. The omitted terms
contain the $\delta$-function, its derivatives, and the
$V\!P$-terms with the poles of the higher orders.

Further on we consider the LO and the NLO approximation only, i.e.
we restrict ourselves by the analysis of the terms displayed by
formula (14). Correspondingly, the cross-section $\hat\sigma(s)$
can be represented as
\begin{equation}\label{15}
\hat\sigma(s) = \hat\sigma_0(s) + \hat\sigma_1(s) ,
\end{equation}
where $\hat\sigma_0(s)$ and $\hat\sigma_1(s)$ stand for the LO and
NLO approximations. Combining (4), (5) and (14), we get
\begin{equation}\label{16}
\hat\sigma_0(s) =\theta(s-4M^2) \times
\frac{1}{s^2}\sqrt{\lambda(s;M^2,M^2}\;F(s;M^2,M^2)\,.
\end{equation}
The $\hat\sigma_1(s)$ can be represented in a form of sum of two
terms, $\hat\sigma_1(s) = \hat\sigma_1^{\mbox{\scriptsize an}}(s)
+ \hat\sigma_1^{V\!P}(s)$. Here the first term, the so-called
``anomalous'' term \cite{N1,N2}, differs on a factor from
$\hat\sigma_0(s)$,
\begin{equation}\label{17}
\hat\sigma_1^{\mbox{\scriptsize an}}(s) =
-2\,\frac{\alpha\Gamma_1}{\Gamma_0}\,\hat\sigma_0(s)\,.
\end{equation}
The second term includes the $V\!P$ contribution,
\begin{eqnarray}\label{18}
\lefteqn{ \hat\sigma_1^{V\!P}(s) = 2\,
\frac{M\Gamma_0}{\pi}\:\theta\!\left(s-(M+M_W)^2\right)
\vspace*{-\baselineskip} }\nonumber
\\
&& \qquad\>\:\times\!\!\int\limits_{M_W^2}^{(\sqrt{s}\!-\!M)^2}
\!\d \tilde{s}\: \frac{1}{s^2}\sqrt{\lambda(s;M^2,\tilde{s})}\>
F(s;M^2,\tilde{s}) \times V\!P\frac{1}{(\tilde{s}-M^2)^2}\,.
\end{eqnarray}
Factor 2 in both formulas, (17) and (18), appears in view of the
pairness of production of unstable particles in the given process.

It is worth noticing that $\hat\sigma_1^{V\!P}(s)$, in contrast to
$\hat\sigma_0(s)$ and $\hat\sigma_1^{\mbox{\scriptsize an}}(s)$,
includes nonzero contribution below the threshold.\footnote{
Nevertheless this contribution becomes zero below $\sqrt{s}=M+M_W$
since in NLO one of two $t$ quarks is produced on-shell. The NNLO
approximation contains contributions with two $t$ quarks produced
off-shell and these contributions are nonzero till the low
physical threshold, $\sqrt{s} = 2M_W$ with the massless $b$
quarks. } The principal-value symbol in (18) at its calculating
can be omitted because the double-pole singularity turns out to be
beyond the integrating area. Correspondingly, at $s \to 4M^2$ from
below, the function $\hat\sigma_{1}^{V\!P}(s)$ possesses a
singular behavior,
\begin{equation}\label{19}
\hat\sigma_{1}^{V\!P}(s)_{\left| \scriptstyle s\,\to\,4M^2 \atop
\scriptstyle s\,<\,4M^2 \right.} = \frac{\Gamma_0}{(2M)^2}\:
F(4M^2,M^2,M^2)\left(4M^2\!-s\right)^{-1/2} .
\end{equation}
At $s>4M^2$ the double pole in (18) gets into the integration area
and is regularized by the principal value. Correspondingly, at $s
\to 4M^2$ from above, integral (18) takes a finite value.

Substituting (16)--(18) into formula (1) and calculating the
convolution, we get the total cross-section $\sigma(s)$ with
taking into consideration the effect of the initial state
radiation. In the case of $\hat\sigma_0(s)$ and
$\hat\sigma_1^{\mbox {\scriptsize an}}(s)$, in so doing, we do not
meet any problems. Below the~threshold their contributions are
zero due to the $\theta$-function in (16), and while going over
the threshold $s = 4M^2$ the nonzero contributions are
switching-off continuously. In the case of
$\hat\sigma_1^{V\!P}(s)$ we have completely different situation
because of singularity (19). Really, though this singularity is
integrable the contribution of $\hat\sigma_1^{V\!P}(s)$ to
$\sigma(s)$ turns out to be divergent at $s \to 4M^2$ (from above
and below) as $|s-4M^2|^{\beta - 1/2}$. Such a behavior is the
consequence of a superposition of singularities in the flux
function $\phi(z;s)$ at $z\to 1$ and in the function
$\hat\sigma_{1}^{V\!P}(zs)$ at $zs \to 4M^2$. The occurrence of
the divergency in $\sigma(s)$ means that its expansion in powers
of $\alpha$ becomes formal and certainly has bad convergence
properties in a neighborhood of the threshold.

Actually, the latter property is a common one for all the
processes with production and decay of unstable particles. The
appearance of the divergencies in the MPT series is a
manifestation of a nonanalyticity in the coupling constant
occurring on the threshold in the cross-section $\sigma(s)$. In
turn, the occurrence of the nonanalyticity is conditioned by the
convolution of the hard-scattering cross-section $\hat\sigma(s)$
with the photon flux function $\phi(z;s)$ which is singular and
includes a nonanalyticity in the missing energy in the limit of
the soft photons (at $z \to 1$). The fact of presence of the
nonanalyticity in the coupling constant in the convoluted
cross-section at the threshold can be verified by direct
calculation, most easily in the case of a single production and
decay of unstable particles.

As mentioned above, a way to solve the problem is to Dyson resum
the MPT series. This camouflages the nonanalyticity in the
coupling constant and the cross-section becomes a regular function
at the threshold. However after the Dyson resummation the property
of the gauge cancellations becomes not apparent. So, it makes
sense to take advantage of another strategy, which is to shift
from the threshold and to not consider this region at all while
considering the convoluted cross-section. In this case all terms
of the complete MPT expansion of $\sigma(s)$ appear to be finite
and it becomes possible to state a problem about the precision of
approximation of the series by a finite number of its terms. The
precision of such an approximation is determined by the first
discarded term. For example, the error of the approximation of
$\sigma(s)$ by $\sigma_0(s)$ can be estimated as $O(\sigma_1(s))$.
More precisely, there is an asymptotic estimate $\sigma(s) =
\sigma_0(s) + O(\sigma_1(s))$ which determines the LO
approximation of $\sigma(s)$. The LO+NLO approximation is
determined by the sum of two terms, $\sigma(s) = \sigma_0(s) +
\sigma_1(s)$, with an error $O(\sigma_2(s))$ which is to be
determined in the framework of NNLO approximation.

The definition of $\hat{\sigma}_2(s)$ which is necessary for
calculating $\sigma_2(s)$ is reduced to the integrals of the above
type and also to the double integral which corresponds to product
of the $V\!P$-terms of formula (14). The calculating of the latter
integral is a nontrivial task since a naive calculation leads to
an occurrence of a nonintegrable singularity in
$\hat{\sigma}_2(s)$ at $s\to 4M^2$. The occurrence of a
nonintegrable singularity means that the function
$\hat{\sigma}_2(s)$ should be considered as a distribution, by the
complete analogy to the case of a single $|\Delta(s)|^2$. The
solution of this problem, however, is beyond the scope of the
present paper. So, in the next Section we carry out the
``phenomenological'' determination of the error $O(\sigma_2(s))$
basing on the direct evaluation of $\sigma(s)$ and
$\sigma_0(s)+\sigma_1(s)$.

\section{Numerical results and discussion}

On the basis of the above consideration we are able to begin the
searching for the ranges of $s$ where the LO and NLO
approximations of MPT operate satisfactorily. First we find $s$
where $\sigma_1(s)$ is small enough. (Below we determine the
meaning of the term ``small enough''.) Thus we determine the range
of satisfactory action of the LO approximation. Then we compare
$\sigma_0(s) + \sigma_1(s)$ with $\sigma(s)$ calculated by means
of formulas of Section~2 and find the values of $s$ where
$\sigma(s)-[\sigma_0(s)+\sigma_1(s)]$ is small enough. This will
be the ``phenomenological'' determination of the range of
applicability of LO+NLO approximation.

We use $M = 175 \: \mbox{GeV}$, $M_W = 80.4 \: \mbox{GeV}$, and
the following input-data for the width \cite{Top}:
\begin{equation}\label{20}
 \Gamma_0 = 1.56 \: \mbox{GeV} \,,\quad
 \Gamma_0+\alpha\Gamma_1 = 1.45 \: \mbox{GeV}\,.
\end{equation}
Function $F(s;s_+,s_-)$ introduced by formula (4) we calculate by
means of CompHEP package \cite {CompHEP} with the subsequent
phase-space integration as described in \cite{GENTLE}. The results
of calculations of $\sigma(s)$, $\sigma_0(s)$ and
$\sigma_0(s)+\sigma_1(s)$ are represented in the vicinity of the
threshold by Fig.2a and above the threshold by Fig.3a. The results
for the DPA cross-section are represented by the same figures,
where $\sigma_{\mbox{\scriptsize DPA}}(s)$ is determined by the
same formulas that were used for $\sigma(s)$, but with
substituting $F(s;M^2,M^2)$ for $F(s;s_+,s_-)$ and $|s-M^2 + \i M
\Gamma|^{-2}$ for $|\Delta(s)|^2$. As the ``exact'' $\Gamma$ we
use $\Gamma_0+\alpha\Gamma_1+\alpha^2\Gamma_2 = 1.42\: \mbox{GeV}$
\cite{Top}.\footnote{ Unfortunately, $\alpha^2\Gamma_2$ is
determined not finally because only the QCD two-loop corrections
have been taken into consideration. This fact introduces some
conditionality into the subsequent discussion.} The $\sigma_0(s)$,
$\sigma_0(s)+\sigma_1(s)$ and $\sigma_{\mbox{\scriptsize DPA}}(s)$
in percentage of $\sigma(s)$ are represented by Fig.2b and Fig.3b.
In the same place we represent the results for
$\sigma_1(s)/\sigma(s)$, more precisely for $100\% +
\sigma_1(s)/\sigma(s)[\%]$ where $100\%$ is added for the
convenience of representation of the results on the same figures.

For the analysis of the above results a criterion of smallness of
the corrections must be fixed. For this purpose we again make use
of the input-data (20) where $\Gamma_0$ and $\alpha\Gamma_1$ are
the LO and NLO approximations for the width. In what follows we
consider $\Gamma_0$ and $\alpha\Gamma_1$ by typical
representatives for the LO and NLO approximations and use them for
the definition of the corresponding scales. In so doing, we also
take into consideration the factor 2 that arises in the
cross-section by virtue of the pairness of the production of
unstable particles. So, $\sigma_1(s)$ will be considered small if
the ratio $\sigma_1(s)/\sigma(s)$ will be no more or about
$2\times \alpha \Gamma_1/\Gamma = 15.5\%$. For the correction
$\sigma_2(s)$ the analogous estimate makes up $2\times \alpha^2
\Gamma_2/\Gamma = 4.2\%$. The ``phenomenological'' criteria are as
follows: the LO approximation could deviate from the exact result
by no more or about $2\times |\,\Gamma - \Gamma_0 | / \Gamma =
19.7 \%$. For the LO+NLO approximation the similar estimate makes
up $2\times |\,\Gamma - \Gamma_0 - \alpha \Gamma_1|/\Gamma =
4.2\%$.

By combining the above results we see that the correction
$\sigma_1(s)$ is small enough ($\leq 15.5\%$), beginning
practically with the threshold, more precisely with $\sqrt{s} =
350.8\,\mbox{GeV}$, and persisting up to the end of the tested
area, $\sqrt{s} = 1\,\mbox{TeV}$. In the same region also the
quantity $\sigma(s)-\sigma_0(s)$ is small ($ \leq 19.7 \%$), which
is responsible for the ``phenomenological'' criterion of the
applicability of LO.

As for the LO+NLO approximation, until the correction $\sigma_2$
is calculated we have the ``phenomenological'' criterion only.
According to this criterion the LO+NLO approximation is good in
two narrow regions placed in the neighborhood of the threshold,
and in a broad region $450\,\mbox{GeV} \le \sqrt{s} \le 650
\,\mbox {GeV}$. The first two regions seemingly are not too
reliable in view of the vicinity of the ``cutoff'' of LO and of
the large NLO corrections. The third, broad region satisfies to
all criteria. Moreover, at the larger energies the exceeding of
the LO+NLO approximation of the $4.2\%$-limits is hardly
noticeable. For instance, at the end of the tested area $\sqrt{s}
= 1 \,\mbox{TeV}$ the discrepancy $\sigma (s) -
[\sigma_0(s)+\sigma_1(s)] $ makes up only $6.3\%$ of $\sigma(s)$,
which practically falls under the definition ``about'' $4.2\%$.
With this stipulation it is possible to say that the LO+NLO
approximation operates satisfactorily from $450\,\mbox{GeV}$ up to
the end of the tested area.

In the practically interesting region for the linear colliders, at
$\sqrt{s} = 500 \,\mbox{GeV}$, the LO+NLO approximation deviates
from the ``exact'' $\sigma(s)$ on $0.2\%$ only. This means that
the MPT approach works very well in this region. For comparison,
the deviation of DPA from the ``exact'' result at $\sqrt{s} = 500
\,\mbox{GeV}$ makes up $6.4\%$, which is more than by factor 10 as
greater. At the other interesting point $\sqrt{s} = 800
\,\mbox{GeV}$ \cite{TESLA} the convergence property of MPT changes
to the worse and the precision of the LO+NLO approximation becomes
$5.5\%$ against $2.9\%$ of DPA. Nevertheless a possibility of the
practical application of MPT in this region is not closed since
the results in MPT can be further improved by means of evaluating
the NNLO correction. For getting greater clearness in this
question also the two-loop correction to the width of $t$ quark is
necessary.

In closing this Section it is worth mentioning that the above
results are found rather stable in a relative calculus with
respect to varying the function $F(s;s_+,s_-)$. This gives a
ground to expect that the similar results can take place in other
processes with the pair production of unstable particles.
Moreover, one can expect that our results will remain in force
with taking into account the loop corrections in the function
$F(s;s_+,s_-)$, including the case when just the loop corrections
are discussed. If this is the case, then the MPT opens an
opportunity to unprecedentedly increase the precision of the
description of the processes with the pair production and decay of
unstable particles at the energies near the maximum of the
cross-section.

\section{Conclusion}

The analysis of this paper shows that the MPT approach considered
with the complete expansion in powers of the coupling constant
(without Dyson resummation) operates satisfactorily in the range
of the maximum of the cross-section and somewhat higher in the
energies. This result is obtained in the case of the pair
production of $t$ quarks in the reaction of $e^+e^-$ annihilation
within the LO+NLO approximation, and can be further improved by
attracting the higher orders of the MPT expansion. At the point
$\sqrt{s} = 500 \,\mbox{GeV}$, designed by TESLA and other
linear-collider projects, the LO+NLO approximation of MPT more
than on one decimal order exceeds in accuracy the DPA approach.

\medskip

The author is grateful to D.Bardin for useful discussions and the
interest to this work.

\newpage

\begin{figure}[t]
\[
\epsfbox{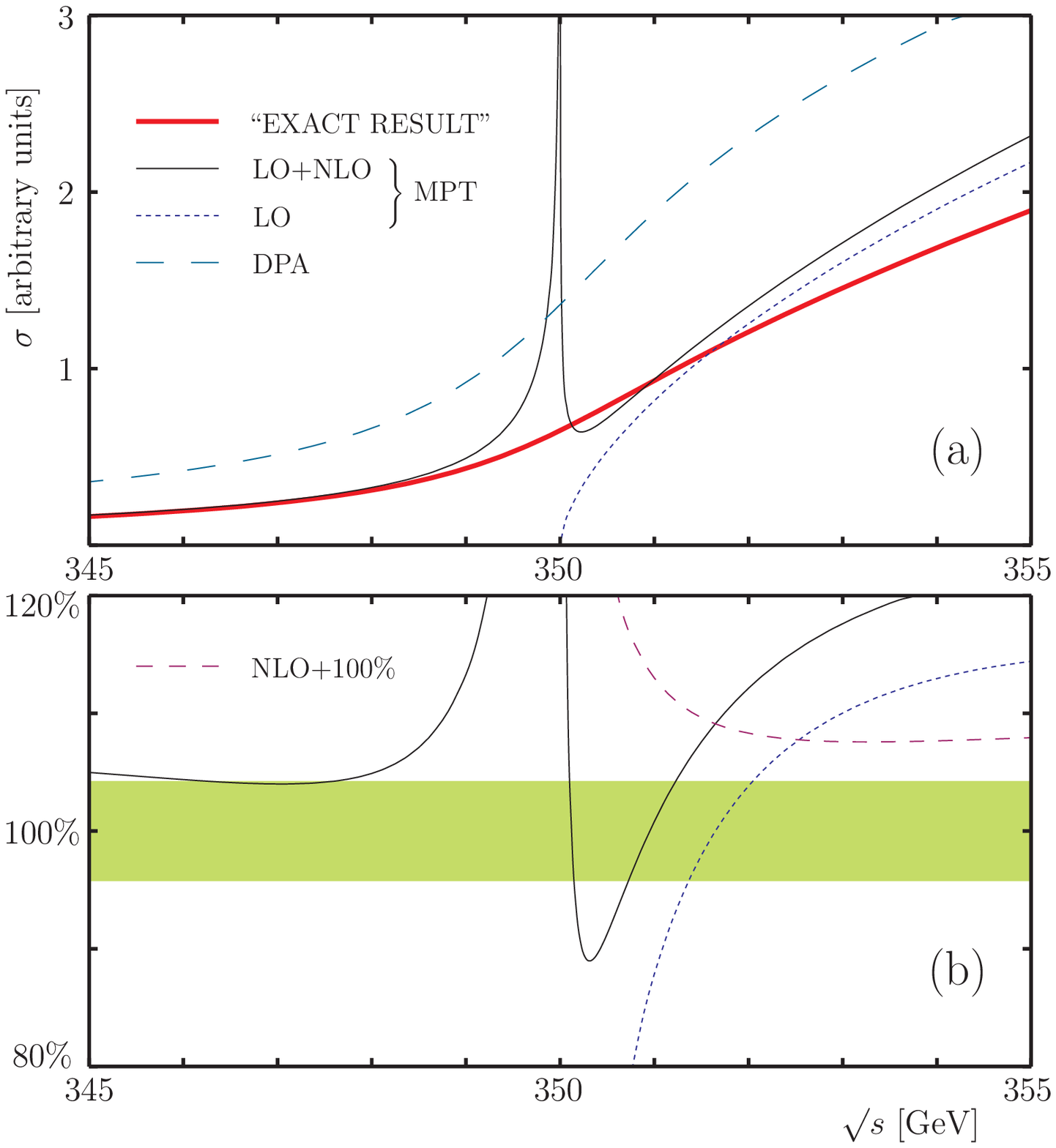}
\]
\caption{The behavior of $\sigma(s)$, $\sigma_0(s)$, $\sigma_0(s)
+ \sigma_1(s)$ and $\sigma_{\mbox{\scriptsize DPA}}(s)$ in a
neighborhood of the threshold. In Fig.2a the results are presented
in absolute arbitrary units, in Fig.2b in percentage of
$\sigma(s)$. In Fig.2b also the results for $\sigma_1(s)$ in
percentage of $\sigma(s)$ are presented with adding $100\%$ for
the convenience of representation of the results on the same
graph. The filled area designates a $4.2\%$-corridor of the
applicability of the LO+NLO approximation.}
\end{figure}
\begin{figure}[t]
\[
\epsfbox{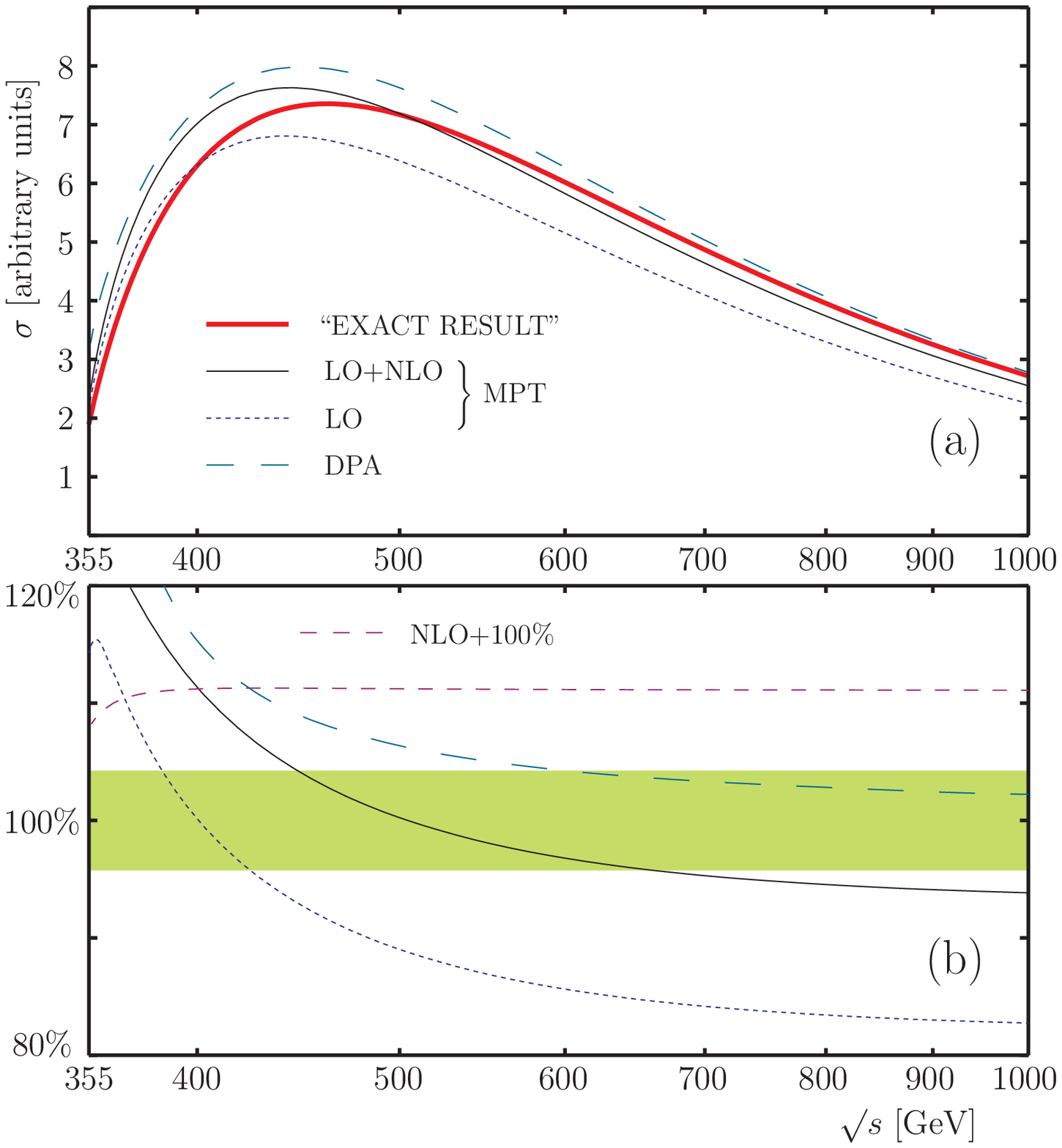}
\]
\caption{The behavior of $\sigma(s)$, $\sigma_0(s)$, $\sigma_0(s)
+ \sigma_1(s)$ and $\sigma_{\mbox{\scriptsize DPA}}(s)$ above the
threshold. The notation is the same that on Fig.2.}
\end{figure}


\begin{thebibliography}{99}
\bibitem{Pole}
R.G.Stuart, Phys.Lett. \textbf{B262} (1991) 113, \textbf{B272}
(1991) 353
\bibitem{4f}
M.W.Gruenewald, et al., The LEP2 Workshop 1999/2000,
``Four-Fermion Production in Electron-Positron Collisions'',
hep-ph/0005309
\bibitem{LC}
E.Accomando et al., Phys.Rep. \textbf{299} (1998) 1
\bibitem{FT}
F.V.Tkachov, in: {\it Proc. of the 32nd PNPI Winter School on
Nuclear and Particle Physics}, ed.by Ya.I.Azimov, et al.
St.Petersburg, PNPI, 1999. p.166 [hep-ph/9802307]
\bibitem{N1}
M.L.Nekrasov, Eur.Phys.J. \textbf{C19} (2001) 441
\bibitem{N2}
M.L.Nekrasov, in: {\it Proc. of XV International Workshop
QFTHEP'2000}, ed.by M.N.Du\-binin et al., SINP MSU, Moscow, 2000,
p.218 [hep-ph/0102284]
\bibitem{Distr}
L.Schwartz, Theorie des Distributions. I, II, Paris, 1950-51;\\
I.M.Gelfand and G.E.Shilov, Generalized Functions. Academic Press,
1968
\bibitem{B}
D.Yu.Bardin, M.L.Nekrasov, (2000) unpublished
\bibitem{OMS}
A.Denner, Fortschr.Phys. \textbf{41} (1993) 307
\bibitem{Hollik}
A.Freitas, W.Hollik, W.Walter, and G.Weiglein, Phys.Lett
\textbf{B495} (2000) 338;\\ A.Freitas, S.Heinemeyer, W.Hollik,
W.Walter, and G.Weiglein, Nucl.Phys.Proc.Suppl. \textbf{89} (2000)
82; \\ A.Freitas, W.Hollik, W.Walter, and G.Weiglein, Nucl.Phys.
\textbf{B632} (2002) 189

\bibitem{OMS-bar}
M.L.Nekrasov, Plys.Lett. \textbf{B531} (2002) 225
\bibitem{Khoze}
I.Bigi, Y.Dokshitzer, V.Khoze, J.Kuhn, P.Zerwas, Plys.Lett.
\textbf{B181} (1986) 157
\bibitem{Top}
M.Beneke et al., Top Quark Physiscs, Cern 2000-004
[hep-ph/0003033]
\bibitem{CompHEP}
A.Pukhov et.al., hep-ph/9908288
\bibitem{GENTLE}
D.Bardin, and T.Riemann, Nucl.Phys. \textbf{B426} (1996) 3
\bibitem{TESLA}
TESLA Technical Design Report, March 2001, Eds. F.Richard, et al.,
\\ http://tesla.desy.de/new$_-$pages/TDR$_-$CD/start.html
\end{thebibliography}
\end{document}